\documentclass[%
 aip,
rsi,%
 amsmath,amssymb,
 reprint,%
]{revtex4-1}

\usepackage{graphicx}
\usepackage{dcolumn}
\usepackage{bm}
\makeatletter
\usepackage{bm}
\usepackage{graphicx}

\makeatother

\begin{document}

\preprint{AIP/123-QED}

\title{Ultra-low noise magnetic field for quantum gases}

\author{Xiao-Tian Xu}
\author{Zong-Yao Wang}
\author{Rui-Heng Jiao}
\author{Chang-Rui Yi}
\author{Wei Sun}
\affiliation{Shanghai Branch, National Research Center for Physical Sciences at Microscale
and Department of Modern Physics, University of Science and Technology
of China, Shanghai 201315, China}
\affiliation{Chinese Academy of Sciences Center for Excellence: Quantum Information and Quantum Physics,
University of Science and Technology of China, Hefei Anhui 230326,
China}

\author{Shuai Chen}
\email{shuai@ustc.edu.cn}

\affiliation{Shanghai Branch, National Research Center for Physical Sciences at Microscale
and Department of Modern Physics, University of Science and Technology
of China, Shanghai 201315, China}
\affiliation{Chinese Academy of Sciences Center for Excellence: Quantum Information and Quantum Physics,
University of Science and Technology of China, Hefei Anhui 230326,
China}

\date{\today}

\begin{abstract}
Ultra-low noise magnetic field is essential for many branches of scientific research.
Examples include experiments conducted on ultra-cold atoms, quantum simulations, as well as precision measurements.
In ultra-cold atom experiments specifically, a bias magnetic field will be often served as a quantization axis and be applied for Zeeman splitting.
As atomic states are usually sensitive to magnetic fields, a magnetic field characterized by ultra-low noise as well as high stability is typically required for experimentation.
For this study, a bias magnetic field is successfully stabilized at 14.5G, with the root mean square (RMS) value of the noise reduced to 18.5$\mu$G (1.28ppm) by placing $\mu$-metal magnetic shields together with a dynamical feedback circuit.
Long-time instability is also regulated consistently below 7$\mu$G.
The level of noise exhibited in the bias magnetic field is further confirmed by evaluating the coherence time of a Bose-Einstein condensate characterized by Rabi oscillation.
It is concluded that this approach can be applied to other physical systems as well.
\end{abstract}

\pacs{PACS}
\maketitle

\begin{quotation}
\end{quotation}

\subsection{Introduction}

Ultra-cold quantum gases can be utilized for quantum simulations that are both extremely clean and controllable, as well as for precision measurements.
Examples therein include optical lattice-based Bose/Fermi-Hubbard models~\cite{Many body physics,Bose-Hubbard Model,Fermi-Hubbard Model,P.Zoller,phase transition,Fermi-Hubbard antiferromagnet}, artificial gauge fields~\cite{gauge field,Zhai,Gauge Potentials}, Feshbach resonances~\cite{Feshbach resonance,Feshbach resonance-2,Feshbach resonance-3}, and optical clocks~\cite{optical clock,NIST}.
However, the environmental noise that occurs in ultra-cold atoms will cause decoherence in systems, presenting a particular drawback when utilizing such ideal experimental platforms.
Such decoherence will influence the accuracy of observations in experimentation adversely, in addition to reducing coherent time in the dynamic evolution of systems.
In many ultra-cold atom experiments, a static bias magnetic field will be used to provide for a quantization axis, for Zeeman effect, or for a Feshbach resonance field (typically between a few and a few hundred Gauss).
The noise exhibited in a given magnetic field will most often be the dominant factor in the environmental noises.
The capability to reduce magnetic field noise empowers researchers to offer much cleaner experimental environments.
This in turn enables researchers to conduct experiments for exotic quantum phenomena and ultra-high accuracy measurements at considerably higher levels of sophistication.

The typical sources of magnetic field noise can be quite diverse.
The Earth's magnetic field will exhibit a daily fluctuation of several hundred micro-Gauss~\cite{Magnetic Field Fluctuations,Geomagnetism}.
The 50 or 60Hz signals that are emitted from the world's power grids, as well as their harmonics, will produce a magnetic field noise that covers a range of values from sub milli-Gauss up to tens of milli-Gauss.
Common lab instruments will also cause similar environmental noise.
The current running through coils will typically generate a static bias magnetic field that will generate magnetic noises as well.
To compensate or shield against such noise will most often demand great effort on the part of researchers.
In most cases, such efforts will be hampered or limited by the design of a given experimental apparatus.
Moreover, it is particularly challenging to suppress magnetic field noise down to the level of tens of micro-Gauss, a level that is required for most high precision experiments~\cite{Spin squeezing,Site-resolved measurement,Microscopy,spin-mix,Li You,wu2016realization,huang2016experimental} characterized by magnetically sensitive internal states.

In experiments, people have found various methods for reducing noise from magnetic fields.
Magnetic shielding is typically adopted to reduce the noise of stray magnetic fields in thermal atom experiments~\cite{hot atoms-1,hot atoms-2}.
For ultra-cold atom experiments, it is oftentimes difficult to shield an entire system, due to the inherent complexities in optical and electrical model designs therein.
In order to achieve a low noise magnetic field, dynamic feedback will be utilized~\cite{ultra-cold atoms-1,ultra-cold atoms-2}, most often while synchronizing experiment circles to 50 or 60Hz signals emanating from a power line.
Regardless, problems persist regarding long time field drifts.
In past research on spin squeezing system~\cite{Diploma thesis}, the noise of a magnetic field was reduced to about 100$\mu$G using dynamic feedback.
For one experiment conducted on an $^{87}$Rb Bose-Einstein condensate (BEC) system~\cite{cold atom-1}, magnetic shielding was used to achieve magnetic field stability that is greater than 50$\mu$G.
In another BEC system experiment~\cite{cold atom-2}, the stray alternating current field strength was reduced to about 40$\mu$G by using magnetic shielding in conjunction with dynamic feedback.
More recently, it was reported that a very low noise magnetic field was generated for an experiment on trapped ion system.
By combining feed-forward and dynamic feedback techniques, a reduced environmental noise of 43$\mu$G, with the bias magnetic field of 146G, has also been achieved~\cite{arXiv}.

\begin{figure*}
\includegraphics[width=0.80\linewidth]{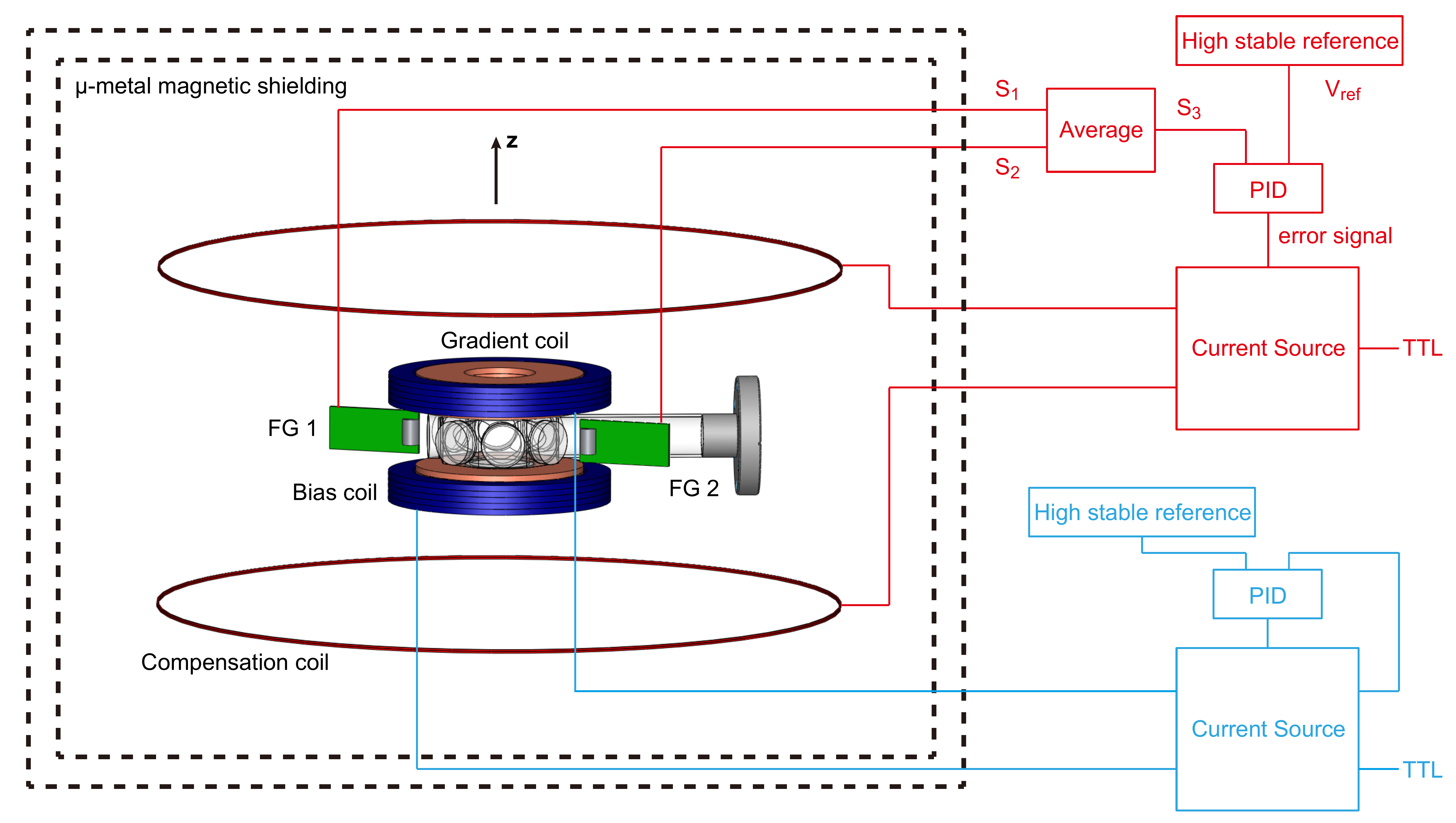}
\caption{\label{Scheme}A schematic of the study model's $\mu$-metal magnetic shields (dashed lines) as well as its feedback loop (solid lines).
The two dashed rectangles illustrate how the $\mu$-metal magnetic shield is double layered.
The color blue denotes the home-built current source for the bias magnetic field, which drives the bias coils.
The color red denotes the components of the dynamic feedback loop used here to stabilize the bias magnetic field.
The fluxgate magnetic field sensors (FG 1 and FG 2) were installed symmetrically and outside the octagon glass chamber, in order to monitor the bias magnetic field.
The model's power supply, which drives its compensation coils, runs in a constant-voltage mode.
The TTL signals serve as switches in this model.}
\end{figure*}

For this study, magnetic shielding as well as dynamic feedback are utilized together in order to suppress magnetic field noise as well as increase stability in a bias magnetic field.
This is achieved through the study model's compact construction for ultra-cold atom experiment.
It is found that the environmental noise from the magnetic field generated in the study's experiment region is reduced below 30$\mu$G using two layers of $\mu$-metal magnetic shielding.
The 14.5G bias magnetic field generated in the experiment is further stabilized using a dynamic sampling-feedback approach, reducing RMS value of noise to 18.5$\mu$G.
Long-time drift is also decreased below 7$\mu$G within two hours.
It is found that the coherence time for an $^{87}$Rb BEC between $|1,-1\rangle$ and $|1,0\rangle$ states are extended to $11.6 \text{ms}$, a result that is consist with the field noise generated in the experiment.
These results are found to be one or two orders of magnitude longer when compared with past experiments in the field~\cite{wu2016realization,Zhang,NP-Ji,PRL-Ji,new2D SOC-W.Sun}.
Similar low noise magnetic fields have already applied in recent experiments to precisely map out the band structures as well as observe the post-quench dynamic processes in cold atoms~\cite{Xu, quench-dynamics}.
This study's results present additional research opportunities to investigate the interesting physics involved in long periods of quantum coherence.

\subsection{Experimental setup}

The experimental setup is illustrated in Fig.1.
The octagonal glass chamber in the center is the science chamber used in performing the study's ultra-cold quantum gas experiment.
The cold atoms are trapped and then cooled to condensation in the center of the chamber.
Two pairs of coils are installed just outside the glass chamber to provide a magnetic field for the experiment.
The inner coils, in an anti-Helmholtz configuration, work to generate a gradient magnetic field for the model's magneto-optical trap as well as quadrupole magnetic trap.
The outer coils, in a Helmholtz configuration, provide a uniform bias magnetic field along z direction in order to generate the Zeeman splitting.
A third set of compensation coils (used here to reduce noise and stabilize the bias magnetic field) are located symmetrically, just beyond the aforementioned two pairs of coils.
The entire vacuum system, together with the optics surrounding it, are enclosed by two layers of $\mu$-metal magnetic shielding, with the combined dimensions of $1.3\text{m} \times 1.0\text{m} \times 1.0\text{m}$.
Two fluxgate magnetic field sensors (Stefan Mayer FL1-1000) are placed symmetrically just outside the octagonal glass chamber in order to make signal observations and monitor fluctuations in the magnetic field within the study's working area.

The model's bias coils are driven by a homemade, low noise current source that operates on 0 to 10A.
The bias coils can establish a bias magnetic field from 0 to 30Gauss precisely.
The low noise current source, also homemade, is regulated by a voltage reference with high stability as well as a proportional-integral-differential (PID) controller (Stanford Research Systems SIM960).
A transistor-transistor logic (TTL) signal is used to switch the current in the coils on or off.
The bias magnetic field is stabilized by utilizing a dynamic sampling-feedback approach, and is measured using two fluxgate sensors.
The signals from these two sensors ($S_{1}$ and $S_{2}$) are averaged as the signal $S_{3}$.
The value of the latter will indicate the properties of the magnetic field at the center of the glass chamber as a result of the study model's symmetrical construction.
The average signal $S_{3}$ was compared with the fixed ultra-low noise voltage $V_{ref}$, as provided by the high stability voltage reference (Stanford Research Systems SIM960).
An error signal is sent to the PID controller to drive the model's low noise power supply (Agilent 6612C).
This is done in order to adjust the current in the compensation coils so as to stabilize the bias magnetic field.

\begin{figure*}
\includegraphics[width=0.90\linewidth]{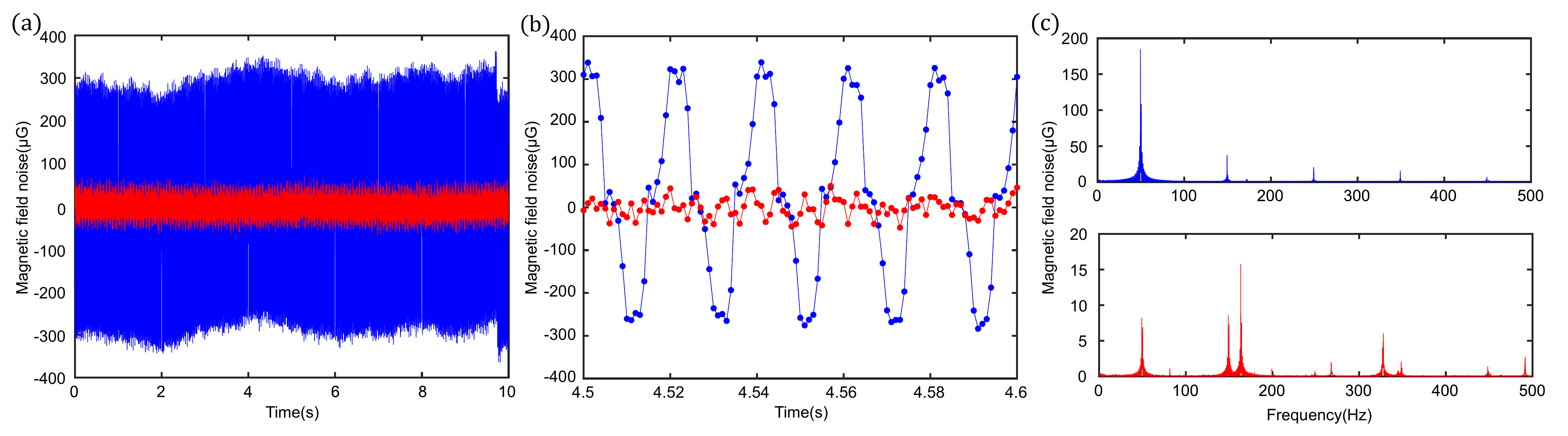}
\caption{\label{noise1}
(a) The blue parts of the spectrum denote the environmental magnetic field noise that occurs outside of the $\mu$-metal magnetic shielding, while the red parts denote the noise that occurs within the shields, over time.
Specifically, the RMS value for the environmental magnetic field noise is 200.3$\mu$G outside of the shielding, while the value is 23.6$\mu$G for inside the shields.
(b) The data of the noise, as taken from (a), from 4.5s to 4.6s specifically.
(c) The upper pane illustrates, in the frequency domain, the environmental magnetic field noise outside the $\mu$-metal magnetic shields, while the lower pane depicts noise that occurs within the shields.
The main component of the environmental magnetic field noise is at 50Hz and its odd harmonics.
}
\end{figure*}

\begin{figure*}
\includegraphics[width=0.90\linewidth]{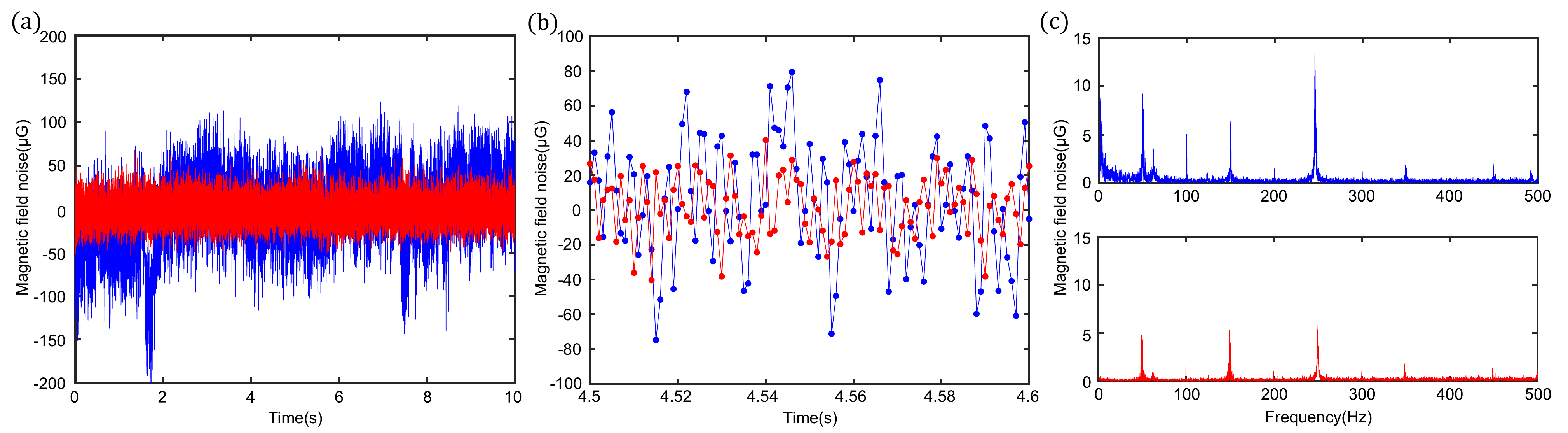}
\caption{\label{noise2}
(a) The blue parts of the spectrum denote the bias magnetic field noise that occurs inside the $\mu$-metal magnetic shields in the time domain before stabilization, while the red parts denote the noise after stabilization.
The RMS value for the bias magnetic field noise is 44.3$\mu$G before stabilization, and 18.5$\mu$G after stabilization.
(b) The data of the noise, as taken from (a), from 4.5s to 4.6s specifically.
(c) The upper pane illustrates, in the frequency domain, the bias magnetic field noise that occurs inside the $\mu$-metal magnetic shields before stabilization, while the lower pane depicts the noise that occurs after stabilization.
}
\end{figure*}

\subsection{Results}

The primary contributions to noise in the magnetic field during research and experiment are found in the lab environment itself.
Examples include power lines as well as any the other electrical instruments in the immediate vicinity.
Without magnetic shields, the magnetic field along z direction in this study was 220mG, as measured by the fluxgate sensors FG 1 and FG 2.
An illustration of the noise therein is presented in Fig.2(a) and Fig.2(b), highlighted in blue.
The data therein was taken over 10 seconds, as shown in Fig.2(a), with the time scale from 4.5s to 4.6s presented in Fig.2(b) for greater detail.
The peak-to-peak value for the signal highlighted in blue in Fig.2(a) was found to be approximately 650$\mu$G, with an RMS value of 200.3$\mu$G.
A fast Fourier transform (FFT) was applied in order to analyze the frequency components of the observed noise, as shown in Fig.2(c).
It was found that the primary contributors were a 50Hz signal and its corresponding odd harmonics, as generated from the study lab's power line.
Long-term drift measured around several milli-Gauss over 2 hours.

The first step in suppressing environment noise in the magnetic field was to enclose the experiment setup within two layers of $\mu$-metal magnetic shielding.
After shielding, it was found that the residual magnetic field along z direction decreased to about 0.5mG.
The noise was also greatly reduced as a result.
This is highlighted in red in Fig.2(a) and Fig.2(b).
Subsequently, the peak-to-peak value of the field noise was found to be approximately 100$\mu$G, with an RMS value reduced to 23.6$\mu$G.
Another FFT was applied to analyze the frequency components of the noise, with results presented in Fig.2(c).
It was observed that the 50Hz component emanating from the lab's power line was suppressed from 184.6$\mu$G to 8.2$\mu$G, and that the values for the 150Hz component decreased from 37.2$\mu$G to 8.6$\mu$G.
This represents a significant decrease in noise.
The remaining 164Hz frequency component, as well as its secondary harmonics, exhibited an amplitude of 15.7$\mu$G and 6.0$\mu$G, respectively.
However, it must be noted that the exact source of this last component remained unclear.

When the bias magnetic field (as generated by the model's bias coils) was switched on, additional magnetic field noise was generated from the current source.
The bias coils were driven by a homemade, low noise current source.
This current source used batteries for its power supply, and was regulated by the model's high stability voltage reference as well as its PID controller.
The relative noise of the current was found to be $3.0 \times 10^{-6}$.
This was measured directly via the voltage signal that was exhibited from sampling resistance, determined primarily by, as well as limited by, the stability of the sampling resistance and reference voltage.
The bias magnetic field was set at 14.5G for this study's experiment.
A graph of the noise observed herein is presented in Fig.3(a) and Fig.3(b), highlighted in blue.
The peak-to-peak value for the magnetic field noise was found to be approximately 200$\mu$G, with a corresponding RMS value of 44.3$\mu$G.
The noise spectrum obtained using an FFT analysis is shown in Fig.3(c).
The primary frequency components of the bias magnetic field noise was found to be the 50Hz signal and its harmonics.

In order to reduce noise and to stabilize the bias magnetic field further, a dynamic sampling-feedback approach was subsequently utilized.
The bias magnetic field was measured using the two fluxgate magnetic field sensors FG 1 and FG 2, which were placed symmetrically outside the study model's octagonal glass chamber.
Specifically, the fluxgate sensors' bandwidth is 1 kHz.
The signals of these two sensors ($S_{1}$ and $S_{2}$) are averaged to provide the signal $S_{3}$.
The value of the latter will indicate the properties of the magnetic field at the center of the glass chamber as a result of the study model's symmetrical construction.
The average signal $S_{3}$ was compared with the fixed ultra-low noise voltage $V_{ref}$, as provided by the high stability voltage reference, located inside the PID controller.
Then the PID controller gives the error signal to drive a commercial low noise power supply for adjusting the current in the compensation coils.
By fine tuning the parameters of the PID controller, the bias magnetic field was stabilized further.
The bandwidth of the entire dynamic feedback loop is about 500Hz, and is limited by the inductance in the compensate coils as well as the bandwidth of the fluxgate sensors.
This was already sufficient for suppressing noise.
The residue noise exhibited in the bias magnetic field is presented in Fig.3(a) and Fig.3(b), highlighted in red.
The peak-to-peak value for this in Fig.3(a) was found to be approximately 80$\mu$G, with an RMS value of 18.5$\mu$G.
The results of another FFT analysis are presented in Fig.3(c).
It can be seen that the components of the 50Hz noise, as well as its corresponding harmonics, have been reduced significantly.

\begin{figure}
\includegraphics[width=\linewidth]{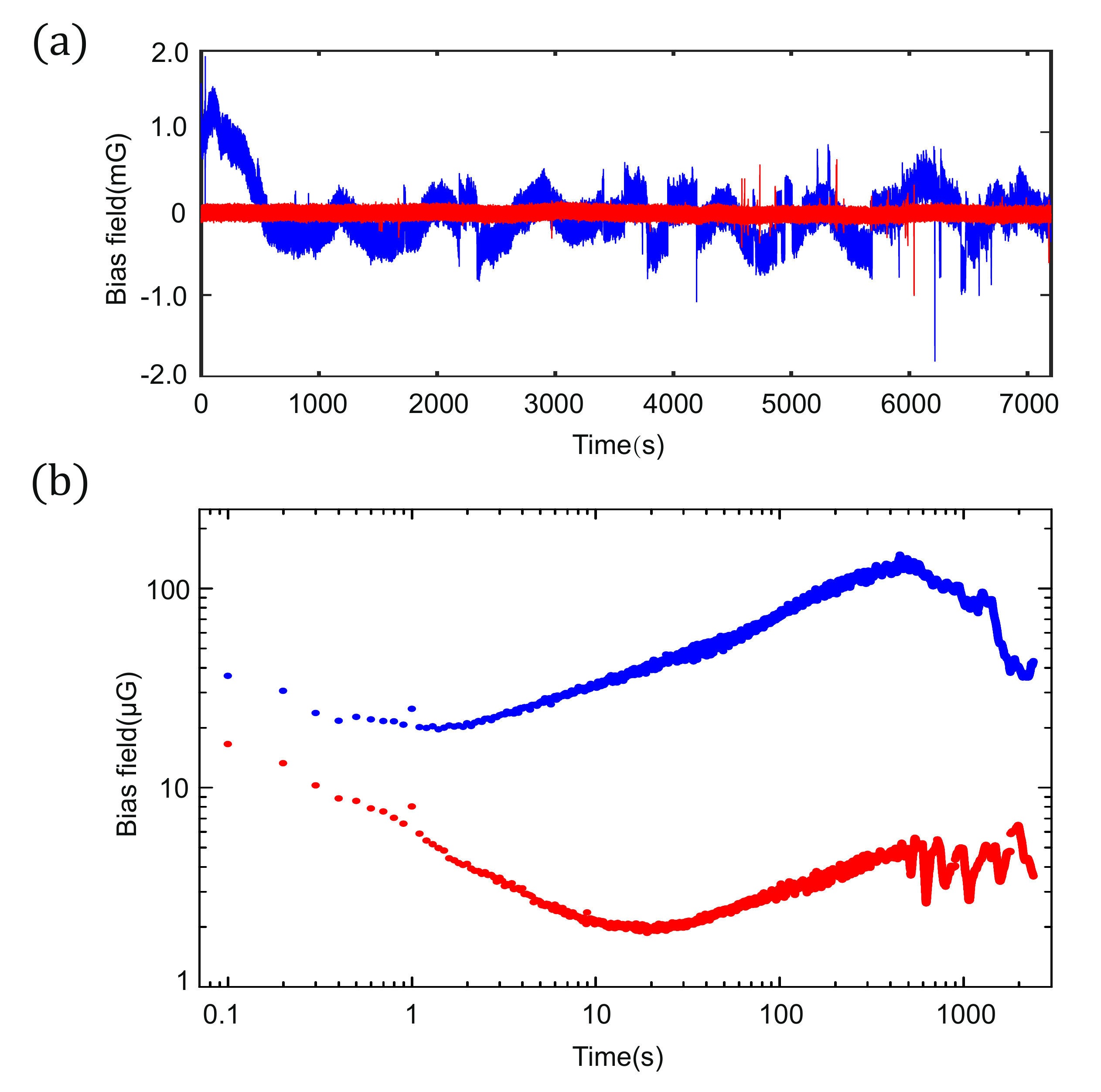}
\caption{\label{stability}The long-term stability observed in the bias magnetic field.
(a) The data of the bias magnetic field, as collected over 2 hours.
The blue lines present the results in free running mode, while the red lines present the results during dynamic feedback.
(b) The results for Allan deviation found in the magnetic field noise.
The blue dots present the results in free running mode, while the red dots present the results during dynamic feedback.
}
\end{figure}

Subsequently, an investigation was conducted into the long-term drift exhibited in the study's bias magnetic field.
With the bias current running inside the $\mu$-metal magnetic shields in free running mode, and with dynamic feedback as well, the bias magnetic field was monitored continuously over two hours.
The observed fluctuations and drifts are presented in Fig.4(a), highlighted in blue and red, respectively.
It can be seen that fluctuations are 1mG in free running mode.
This fluctuation is reduced by over an order of magnitude when dynamic feedback is considered.
An analysis of the exhibited Allan deviation is presented in Fig.4(b).
Here, the short-term to long-term stability of the bias magnetic field can be better understood.
In free running mode, short-term stability is of 36.3$\mu$G on a scale of 0.1s.
The minimum value found therein is about 20$\mu$G, on a scale of 1s.
Stability maintains itself at a level of 100$\mu$G for a long time.
When considering dynamic feedback as well, short-term stability is of 16.5$\mu$G on a scale of 0.1s.
The minimum value of stability found therein reaches 2.1$\mu$G, which is the limit resolution of the fluxgate sensor, on a scale of 20s.
For a period up to two hours, stability in the study system's bias magnetic field is generally below 7$\mu$G.
Note that this value is still smaller than the RMS value observed for the system's noise, which is negligible regardless in the experiment.

\subsection{The Rabi oscillation in a Bose-Einstein condensate in magnetically sensitive states}

Suppression in magnetic field noise was confirmed further by measuring the Rabi oscillation in an $^{87}$Rb BEC.
To clarify, the Rabi oscillation is the oscillation in level populations (or quantum mechanical probability amplitudes) under the influence of an incident light field.
For this experiment, it specifically represents the oscillation in the population of the sample's internal state by microwave transition.
The condensate is initially prepared in an optical trap, with $3 \times 10^{5}$ atoms.
Thereafter, the atoms are pumped into the magnetic sub-level $|1,-1\rangle$.
With the bias magnetic field $B=14.5$G, the $10.216$MHz on resonance Radio-Frequency (RF) pulse with different duration time is applied to couple the state $|1,-1\rangle$ and the state $|1,0\rangle$ in $F=1$ manifold.
The $|1,1\rangle$ state is effectively suppressed due to the relatively large quadratic Zeeman shift.
At the end of the RF pulse, all the lasers are switched off in $1\mu$s and the spin-resolved time-of-flight (TOF) image is taken after $25$ms to analyze the spin distribution in the momentum space.

The Rabi oscillation in the BECs driven by the RF pulse is presented in Fig.5(a).
Here we define the state $|1,-1\rangle$ (yellow circle) as the spin-up state and the state $|1,0\rangle$ (orange circle) as the spin-down state.
A count is subsequently made for the atoms of each state.
The number of atoms in spin-up (spin-down) state is recorded as $N_{\uparrow}$ ($N_{\downarrow}$).
Here, spin-polarization is defined as $P=\frac{N_{\uparrow}-N_{\downarrow}}{N_{\uparrow}+N_{\downarrow}}$.
The evolution of the spin-polarization $P$, over time, is illustrated in Fig.5($\rm b_{1}$).
It can be seen that the spin-polarization $P$ oscillates between -1 and 1.
Due to the decoherence induced by the noise of the magnetic field, the error bar $\Delta P$ becomes larger as time increases.
Then we use the function of $P(t)=e^{-\frac{t}{\tau_{0}}}\cos(\frac{2\pi t}{T_0})$ to fit the data, where $\tau_0$ is the decay time and $T_0$ is the period of time for the Rabi oscillation.
The fitting curve is shown in Fig.5($\rm b_{1}$).
The fitting period $T_0$ of the Rabi oscillation is $0.9225(3)$ms.

\begin{figure}
\includegraphics[width=\linewidth]{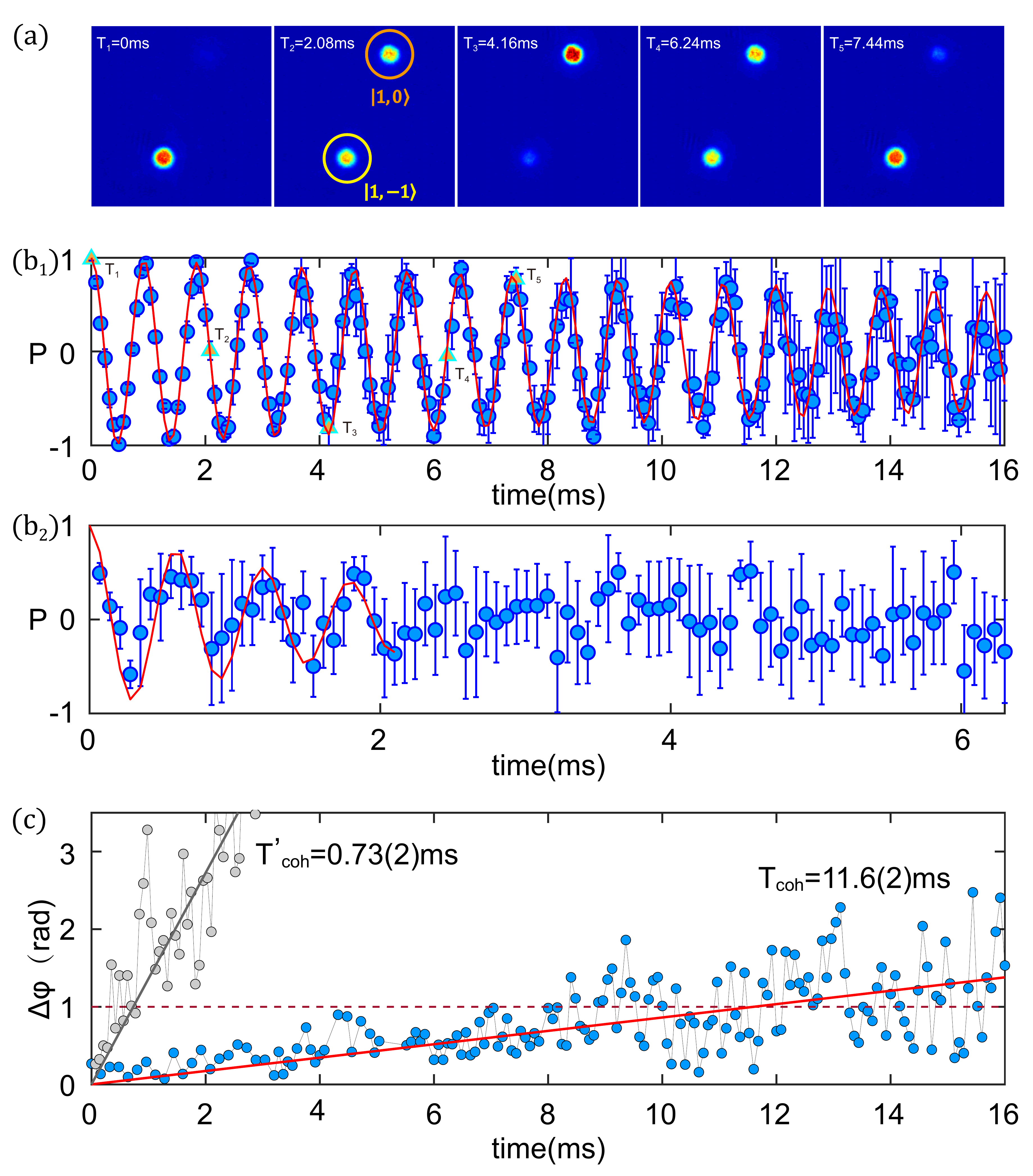}
\caption{\label{Rabi oscillation}The Rabi oscillation observed in the BEC.
(a) Raw oscillation observations taken from various periods in time.
The yellow circle denotes the spin-up state $|1,-1\rangle$, while the orange circle denotes the spin-down state $|1,0\rangle$.
($\rm b_{1}$) The evolution of spin-polarization $P$, over time, and with the magnetic field stabilized.
Note that the fitting period for the Rabi oscillation is $0.9225(3)$ms.
($\rm b_{2}$) The evolution of spin-polarization $P$, over time, but without $\mu$-metal magnetic shielding or dynamic feedback.
Note that the fitting period for the Rabi oscillation is $0.599(8)$ms.
(c) The evolution of phase error $\Delta\varphi$ observed in the Rabi oscillation.
The blue circles indicate results taken from when the magnetic field was stabilized.
The gray circles indicate results taken when there was no $\mu$-metal magnetic shielding or dynamic feedback.
With the magnetic field stabilized, the fitting coherent time is $11.6(2)$ms.
Without $\mu$-metal magnetic shielding or dynamic feedback, the fitting coherent time is $0.73(2)$ms.
}
\end{figure}

People usually study the damping of Rabi oscillation in order to obtain the coherence of two-level quantum system~\cite{coherence-time-1,coherence-time-2,coherence-time-3}.
However, the amplitude of Rabi oscillation decreases very slowly in our experiment.
The residue noise of the magnetic field leads to the diffusion of the phase during Rabi oscillation, cause the increasing error bar with the evolution time increases as shown in Fig.5($\rm b_{1}$). 
The noise can be extracted from the coherence time of the system.
In the ideal conditions (i.e. those lacking any decay or dephase), a Rabi oscillation can be expressed as $P(t)=\cos(\varphi(t))$.
Here $P(t)$ represents the spin-polarization, while $\varphi(t)$ equals to $\frac{2 \pi t}{T}$ under the period of the Rabi oscillation $T$.
In another way, Rabi oscillation can be understood as a point, rotating clockwise, along a unit circle.
The spin-polarization $P(t)$ can be regarded as the point projected on the y-axis.
The phase of the Rabi oscillation $\varphi(t)$ can be found by the angle between the y-axis and the connection to the center of the circle.
When considering about the noise in the magnetic field, the phase of the Rabi oscillation will exhibit perturbation.
Finally, the perturbation will reflect on the error of the spin-polarization $P$.
With a simple geometric relationship, we can easily get the formulas below,
\begin{eqnarray}
&\text{if} \quad  \varphi=n\pi, 1-\cos(\frac{\Delta\varphi}{2})=\Delta P, \nonumber \\
&\text{if} \quad  \varphi\neq n\pi, \mid \cos(\varphi-\frac{\Delta\varphi}{2})-\cos(\varphi+\frac{\Delta\varphi}{2}) \mid=2\Delta P, \nonumber
\end{eqnarray}
where $\Delta P$ is the error bar for the spin-polarization $P$.
The phase error $\Delta\varphi$, taken at various points in time, is presented in Fig.5(c).
These data have been linearly fitted.
The fitting line represents the trend of evolution in the phase error $\Delta\varphi$.
When $\Delta\varphi \geqslant 1$, the Rabi oscillation can be treated as being out of phase.
Thus, we can get the coherent time of $T_{coh}=11.6(2)$ms by the intersection of the fitting line and the line $\Delta\varphi=1$.
The noise in the bias magnetic field is calibrated to be $\delta B=\frac{1}{2\pi T_{coh} \times 0.7\text{MHz/G}}=19.6(3)\mu \text{G}$.
The result agrees well with the measurement in electronic devices.

For comparison, the Rabi oscillation of an $^{87}$Rb BEC is also measured, but without $\mu$-metal magnetic shielding or dynamic feedback.
The results are shown in Fig.5($\rm b_{2}$) and Fig.5(c).
With the same analysis method, we can get the coherent time $T^{'}_{coh}=0.73(2)$ms of the system in the condition that without $\mu$-metal magnetic shielding or dynamic feedback.
It also proves the validity of our method for stabilizing the magnetic field.

\subsection{Conclusion}

In conclusion, we have applied the techniques of magnetic shielding and dynamic feedback to achieve an ultra-low noise and stable bias magnetic field for ultra-cold quantum gases experiments.
The RMS value of the noise has been reduced to 18.5$\mu$G (1.28ppm) in a bias magnetic field of 14.5G.
Here the $\mu$-metal magnetic shielding is essentially important to suppress the environmental magnetic field noise, especially the components of the noise at $50$Hz and its harmonics.
The dynamical feedback loop not only effectively reduces the noise, but also decreases the long-term instability below 7$\mu$G, in the order of $10^{-7}$ according to the bias magnetic field.
We have performed the Rabi oscillation of an $^{87}$Rb BEC in magnetically sensitive states to calibrate the noise.
The coherence time of the system is achieved to be $11.6(2)$ms and the corresponding magnetic field noise is of $19.6(3)\mu$G, which is consist with the RMS noise we measured.
The magnetic shielding plus dynamic feedback is a general approach.
It can be conveniently extended to other physical systems.

\section*{Acknowledgment}

This work was supported by the National Key R\&D Program of China (2016YFA0301601), National Natural Science Foundation of China (N0. 11674301), the Anhui Initiative in Quantum Information Technologies (AHY120000) and the Chinese Academy of Sciences.


\begin{thebibliography}{10}

\bibitem{Many body physics} I. Bloch, J. Dalibard, and W. Zwerger, \emph{Many-body physics with ultracold gases}, Rev. Mod. Phys. {\bf 80}, 885 (2008).

\bibitem{Bose-Hubbard Model} J. M. Zhang and R. X. Dong, \emph{Exact diagonalization: the Bose-Hubbard model as an example}, Eur. J. Phys. {\bf 31}, 3 (2010)

\bibitem{Fermi-Hubbard Model} T. Hensgens, T. Fujita, L. Janssen, X. Li, C. J. Van Diepen, C. Reichl, W. Wegscheider, S. Das Sarma, and L. M. K. Vandersypen, \emph{Quantum simulation of a Fermi-Hubbard model using a semiconductor quantum dot array}, Nature {\bf 548}, 70 (2017).

\bibitem{P.Zoller} D. Jaksch, C. Bruder, J. I. Cirac, C. W. Gardiner, and P. Zoller, \emph{Cold Bosonic Atoms in Optical Lattices}, Phys. Rev. Lett. {\bf 81}, 3108 (1998).

\bibitem{phase transition} M. Greiner, O. Mandel, T. Esslinger, T. W. H\"{a}nsch, and I. Bloch, \emph{Quantum phase transition from a superfluid to a Mott insulator in a gas of ultracold atoms}, Nature {\bf 415}, 39-44 (2002).

\bibitem{Fermi-Hubbard antiferromagnet} A. Mazurenko, C. S. Chiu, G. Ji, M. F. Parsons, M. Kanasz-Nagy, R. Schmidt, F. Grusdt, E. Demler, D. Greif, and M. Greiner, \emph{A cold-atom Fermi-Hubbard antiferromagnet}, Nature {\bf 545}, 462-466 (2017).

\bibitem{gauge field} J. Dalibard, F. Gerbier, G. Juzeli$\bar{u}$nas, and P. \"{O}hberg, \emph{Colloquium: Artificial gauge potentials for neutral atoms}, Rev. Mod. Phys. {\bf 83}, 1523 (2011).

\bibitem{Zhai} H. Zhai, \emph{Degenerate quantum gases with spin-orbit coupling: a review}, Rep. Prog. Phys. {\bf 78}, 026001 (2015).

\bibitem{Gauge Potentials} K. Osterloh, M. Baig, L. Santos, P. Zoller, and M. Lewenstein, \emph{Cold Atoms in Non-Abelian Gauge Potentials: From the Hofstadter "Moth" to Lattice Gauge Theory}, Phys. Rev. Lett. {\bf 95}, 010403 (2005).

\bibitem{Feshbach resonance} C. Chin, R. Grimm, P. Julienne, and E. Tiesinga, \emph{Feshbach resonances in ultracold gases}, Rev. Mod. Phys. {\bf 82}, 1225 (2010).

\bibitem{Feshbach resonance-2} N. Takemura, S. Trebaol, M. Wouters, M. T. Portella-Oberli, and B. Deveaud, \emph{Polaritonic Feshbach resonance}, Nat. Phys. {\bf 10}, 500 (2014).

\bibitem{Feshbach resonance-3} S. Inouye, M. R. Andrews, J. Stenger, H.-J. Miesner, D. M. Stamper-Kurn, and W. Ketterle, \emph{Observation of Feshbach resonances in a BosešCEinstein condensate}, Nature {\bf 392}, 151-154 (1998).

\bibitem{optical clock} B. J. Bloom, T. L. Nicholson, J. R. Williams, S. L. Campbell, M. Bishof, X. Zhang, W. Zhang, S. L. Bromley, and J. Ye, \emph{An optical lattice clock with accuracy and stability at the $10^{-18}$ level}, Nature {\bf 506}, 71 (2014).

\bibitem{NIST} M. A. Lombardi, T. P. Heavner, and S. R. Jefferts, \emph{NIST Primary Frequency Standards and the Realization of the SI Second}, NCSLI Measure {\bf 2} (4), 74 (2007).

\bibitem{Magnetic Field Fluctuations} N. F. Ness, T. L. Skillman, G. S. Scearce, and J. P. Heppner, \emph{Magnetic Field Fluctuations on the Earth and in Space}, JPSJS {\bf 17}, 27 (1962).

\bibitem{Geomagnetism} Jacobs, J. A. (Ed), \emph{Geomagnetism Volumes 1, 2 and 3}, Academic Press (1987); Jacobs, J. A. (Ed), \emph{Geomagnetism Volumes 4}, Academic Press (1991).

\bibitem{Spin squeezing} P. M. Preiss, R. Ma, M. E. Tai, J. Simon, and M. Greiner, \emph{Quantum gas microscopy with spin, atom-number, and multilayer readout}, Phys. Rev. A {\bf 91}, 041602(R) (2015)

\bibitem{Site-resolved measurement} M. F. Parsons, A. Mazurenko, C. S. Chiu, G. Ji, D. Greif, and M. Greiner, \emph{Site-resolved measurement of the spin-correlation function in the Fermi-Hubbard model}, Science {\bf 353}, 1253-1256 (2016).

\bibitem{Microscopy} M. E. Tai, A. Lukin, M. N. Rispoli, R. Schittko, T. Menke, D. Borgnia, P. M. Preiss, F. Grusdt, A. M. Kaufman, and M. Greiner, \emph{Microscopy of the interacting Harper-Hofstadter model in the two-body limit}, Nature {\bf 546}, 519-523 (2017).

\bibitem{spin-mix} C. Gross, H. Strobel, E. Nicklas, T. Zibold, N. Bar-Gill, G. Kurizki, and M. K. Oberthaler, \emph{Atomic homodyne detection of continuous-variable entangled twin-atom states}, Nature {\bf 480}, 219-223 (2011).

\bibitem{Li You} X.-Y. Luo, Y.-Q. Zou, L.-N. Wu, Q. Liu, M.-F. Han, M. K. Tey, and L. You, \emph{Deterministic entanglement generation from driving through quantum phase transitions}, Science {\bf 355}, 620 (2017).

\bibitem{wu2016realization} Z. Wu, L. Zhang, W. Sun, X.-T. Xu, B.-Z. Wang, S.-C. Ji, Y. Deng, S. Chen, X.-J. Liu, and J.-W. Pan, \emph{Realization of Two-Dimensional Spin-Orbit Coupling for Bose-Einstein Condensates}, Science {\bf 354}, 83 (2016).

\bibitem{huang2016experimental} L. Huang, Z. Meng, P. Wang, P. Peng, S.-L. Zhang, L. Chen, D. Li, Q. Zhou, and J. Zhang, \emph{Experimental Realization of Two-Dimensional Synthetic Spin-Orbit Coupling in Ultracold Fermi Gases}, Nat. Phys. {\bf 12}, 540 (2016).

\bibitem{hot atoms-1} Y. Xiao, M. Klein, M. Hohensee, L. Jiang, D. F. Phillips, M. D. Lukin, and R. L. Walsworth, \emph{Slow Light Beam Splitter}, Phys. Rev. Lett. {\bf 101}, 043601 (2008).

\bibitem{hot atoms-2} I. Novikova, R. L. Walsworth, and Y. Xiao, \emph{Electromagnetically induced transparency-based slow and stored light in warm atoms}, Laser \& Photonics Reviews {\bf 6}, 333–353 (2012).

\bibitem{ultra-cold atoms-1} M. Pr\"{u}fer, P. Kunkel, H. Strobel, S. Lannig, D. Linnemann, C.-M. Schmied, J. Berges, T. Gasenzer, and M. K. Oberthaler, \emph{Observation of universal dynamics in a spinor Bose gas far from equilibrium}, Nature {\bf 563}, 217-220 (2018).

\bibitem{ultra-cold atoms-2} C. S. Chiu, G. Ji, A. Mazurenko, D. Greif, and M. Greiner, \emph{Quantum state engineering of a Hubbard system with ultracold fermions}, Phys. Rev. Lett. {\bf 120}, 243201 (2018).

\bibitem{Diploma thesis} Christian Gro$\beta$, \emph{Spin squeezing and non-linear atom interferometry with Bose-Einstein condensates}, Diploma thesis, University of Heidelberg (2010).

\bibitem{cold atom-1} A. \"{O}ttl, S. Ritter, M. K\"{o}hl, and T. Esslinger, \emph{Hybrid apparatus for Bose-Einstein condensation and cavity quantum electrodynamics: Single atom detection in quantum degenerate gases}, Rev. Sci. Instrum. {\bf 77}, 063118 (2006).

\bibitem{cold atom-2} C. J. Dedman, R. G. Dall, L. J. Byron, and A. G. Truscott, \emph{Active cancellation of stray magnetic fields in a Bose-Einstein condensation experiment}, Rev. Sci. Instrum. {\bf 78}, 024703 (2007).

\bibitem{arXiv} B. Merkel, K. Thirumalai, J. E. Tarlton, V. M. Sch\"{a}fer, C. J. Ballance, T. P. Harty, and D. M. Lucas, \emph{Magnetic field stabilization system for atomic physics experiments}, arXiv:1808.03310v1.

\bibitem{Zhang} J.-Y. Zhang, S.-C. Ji, Z. Chen, L. Zhang, Z.-D. Du, B. Yan, G.-S. Pan, B. Zhao, Y.-J. Deng, H. Zhai, S. Chen, and J.-W. Pan, \emph{Collective Dipole Oscillations of a Spin-Orbit Coupled Bose-Einstein Condensate}, Phys. Rev. Lett. {\bf 109}, 115301 (2012).

\bibitem{NP-Ji} S.-C. Ji, J.-Y. Zhang, L. Zhang, Z.-D. Du, W. Zheng, Y.-J. Deng, H. Zhai, S. Chen, and J.-W. Pan, \emph{Experimental Determination of the Finite-Temperature Phase Diagram of a Spin-Orbit Coupled Bose Gas}, Nat. Phys. {\bf 10}, 314 (2014).

\bibitem{PRL-Ji} S.-C. Ji, L. Zhang, X.-T. Xu, Z. Wu, Y. Deng, S. Chen, and J.-W. Pan, \emph{Softening of Roton and Phonon Modes in a Bose-Einstein Condensate with Spin-Orbit Coupling}, Phys. Rev. Lett. {\bf 114}, 105301 (2015).

\bibitem{new2D SOC-W.Sun} W. Sun, B.-Z. Wang, X.-T. Xu, C.-R. Yi, L. Zhang, Z. Wu, Y. Deng, X.-J. Liu, S. Chen, and J.-W. Pan, \emph{Highly Controllable and Robust 2D Spin-Orbit Coupling for Quantum Gases}, Phys. Rev. Lett. {\bf 121}, 150401 (2018).

\bibitem{quench-dynamics} W. Sun, C.-R. Yi, B.-Z. Wang, W.-W. Zhang, B. C. Sanders, X.-T. Xu, Z.-Y. Wang, J. Schmiedmayer, Y. Deng, X.-J. Liu, S. Chen, and J.-W. Pan, \emph{Uncover Topology by Quantum Quench Dynamics}, Phys. Rev. Lett. {\bf 121}, 250403 (2018).

\bibitem{Xu} X.-T. Xu, C.-R. Yi, B.-Z. Wang, W. Sun, Y. Deng, X.-J. Liu, S. Chen, and J.-W. Pan, \emph{Precision mapping the topological bands of 2D spin-orbit coupling with microwave spin-injection spectroscopy}, Sci. Bull. {\bf 63}, 1464 (2018).

\bibitem{coherence-time-1} A. Zrenner, E. Beham, S. Stufler, F. Findeis, M. Bichler, and G. Abstreiter, \emph{Coherent properties of a two-level system based on a quantum-dot photodiode}, Nature {\bf 418}, 612 (2002).

\bibitem{coherence-time-2} D. M. Meekhof, C. Monroe, B. E. King, W. M. Itano, and D. J. Wineland, \emph{Generation of Nonclassical Motional States of a Trapped Atom}, Phys. Rev. Lett. {\bf 76}, 1796 (1996).

\bibitem{coherence-time-3} R. Bonifacio, S. Olivares, P. Tombesi, and D. Vitali, \emph{Non-dissipative decoherence in Rabi oscillation experiments}, J. Mod. Opt. {\bf 47}, 2199-2211 (2000).

\clearpage


\end{thebibliography}
\end{document}